\documentclass[11pt]{article}

\usepackage{graphicx}
\usepackage{amssymb,amsmath}

\date{}
\begin{document}

\title{On Modified Gravity}
\author{{} \\ \small{Ivan Dimitrijevic${}^1$, Branko Dragovich${}^2$, Jelena Grujic${}^3$ and Zoran Rakic${}^1$}\\ {} \\
\small{${}^1$Faculty of Mathematics, University of Belgrade,
Studentski trg 16,  Belgrade, Serbia}\\ \small{${}^2$Institute of
Physics,\, University of Belgrade,\,  Pregrevica 118,\, 11080
Belgrade,\, Serbia} \\   \small{${}^3$Teachers Training Faculty,
University of Belgrade,  Kraljice Natalije 43, Belgrade, Serbia}}
\maketitle

\maketitle

\abstract{We consider some aspects of nonlocal modified gravity,
where nonlocality is of the type $R \mathcal{F}(\Box) R$. In
particular, using ansatz of the form $\Box R = c R^\gamma,$ we
find a few $R(t)$ solutions for the spatially flat FLRW metric.
There are singular and nonsingular bounce solutions. For late
cosmic time, scalar curvature $R(t)$ is in low regime and scale
factor $a(t)$ is decelerated. $R (t) = 0$ satisfies all equations
when $k = -1$.}

\section{Introduction}

General theory of relativity was founded by Einstein at the end of
1915 and has been successfully verified as modern theory of
gravity for the Solar System. It is done by the Einstein equations
of motion for gravitational field:
$
R_{\mu\nu} - \frac{1}{2} R g_{\mu\nu} = \kappa T_{\mu\nu},
$
which can be derived from the Einstein-Hilbert action $S =
\frac{1}{16 \pi G} \int \sqrt{-g} R d^4x + \int \sqrt{-g}
\mathcal{L}_{mat} d^4x .$

Attempts to modify general relativity started already at its early
times and it was mainly motivated by research of  possible
mathematical generalizations. Recently there has been an intensive
activity in gravity modification, motivated by discovery of
accelerating expansion of the Universe, which has not yet
generally accepted theoretical explanation. If general relativity
is theory of gravity for the Universe as a whole then it has to be
some new kind of matter with negative pressure, dubbed {\it dark
energy}, which is responsible for acceleration. However, general
relativity has not been verified at the cosmic scale (low
curvature regime) and dark energy has not been directly detected.
This situation has motivated a new interest in modification of
general relativity, which should be some kind of its
generalization (for a recent review of various approaches, see
\cite{clifton}, and for renormalizability \cite{modesto}). However
there is not a unique way how to modify general relativity. Among
many approaches there are two of them, which have been much
investigated: 1) $f(R)$ theories of gravity (for a review, see
\cite{faraoni}) and 2) nonlocal gravities (see, e.g.
\cite{nojiri,biswas} and references therein).

In the case of $f(R)$ gravity, the Ricci scalar $R$ in the action
is replaced by a function $f(R)$. This is extensively investigated
for the various forms of function $f(R)$. We have had some
investigation when $f(R) = R \cosh{\frac{\alpha R + \beta}{\gamma
R + \delta}}$ and, after completion of research, the results will
be presented elsewhere.

\label{sec:1}

In the sequel we shall consider some aspects of nonlocal gravity.
Nonlocality means that Lagrangian contains an infinite number of
space-time derivatives, i.e. derivatives up  to an infinitive
order in the form of d'Alembert operator $\Box$. In string theory
nonlocality emerges as a consequence of extendedness of strings.
Since string theory contains gravity as well as other kinds of
interaction and matter, it is natural to expect nonlocality not
only in the matter sector   but also in  geometrical sector of
gravity. On some developments in cosmology with nonlocality in the
matter sector one can see, e.g.,
\cite{arefeva,calcagni,barnaby,koshelev} and references therein.
In the next section we shall discuss a nonlocal modification of
only geometry sector of gravity and its corresponding cosmological
solutions (on nonlocality in both sectors, see \cite{calcagni1}).

\section{On a Nonlocal Modification of Gravity}
\label{sec:2}

Under nonlocal modification of gravity we understand replacement
of the Ricci curvature $R$ in the action  by a suitable function
$F (R, \Box), $ where $\Box = \frac{1}{\sqrt{-g}} \partial_{\mu}
\sqrt{-g} g^{\mu\nu} \partial_{\nu}$.

Inspired by \cite{biswas} (for recent developments, see
\cite{koshelev1,koshelev2}), we consider nonlocal Lagrangian
without matter in the form
\begin{equation} \label{lag:1}
S = \displaystyle \int d^{4}x \sqrt{-g}\Big(\frac{R}{16 \pi G} +
\frac{c}{2}R \mathcal{F}(\Box)R \Big),
\end{equation}
which was proposed in \cite{biswas0}, where  $ \mathcal{F}(\Box)=
\displaystyle \sum_{n =0}^{\infty} f_{n}\Box^{n}$ and $c$ is a
constant. By variation of the Lagrangian (\ref{lag:1}) with
respect to metric $g^{\mu\nu}$ one obtains the equation of motion
for $g_{\mu\nu}$
\begin{align} \label{eom:1}
&(1 + 16 \pi G c \mathcal{F}(\Box)R) G_{\mu\nu} =
 4 \pi G c \sum_{n=1}^{+\infty} f_{n}\sum_{l=0}^{n-1} (\partial_{\mu}\Box^{l}R \partial_{\nu}\Box^{n-1-l}R \nonumber \\ &+
\partial_{\nu}\Box^{l}R \partial_{\mu}\Box^{n-1-l}R -
g_{\mu\nu}( g^{\rho\sigma} \partial_{\rho}\Box^{l}R
\partial_{\sigma}\Box^{n-1-l}R + \Box^{l}R \Box^{n-l}R)) \nonumber \\ &-   4
\pi G g_{\mu\nu}c R \mathcal{F}(\Box)R + 16 \pi G c (
D_{\mu}\partial_{\nu} - g_{\mu\nu}\Box)\mathcal{F}(\Box)R .
\end{align}
The trace of (\ref{eom:1}) is also a useful formula and it is

\begin{equation} \label{trace:1}
 \sum_{n=1}^{+\infty} f_{n} \sum_{l=0}^{n-1} (\partial_{\mu}\Box^{l}R \partial^{\mu}\Box^{n-1-l}R + 2\Box^{l}R \Box^{n-l}R )
 + 6 \Box \mathcal{F}(\Box) R = \frac{R}{8 \pi G c}.
\end{equation}

We mainly use the spatially flat (homogeneous and isotropic)
Friedmann-Lema\^{\i}tre-Robertson-Walker (FLRW) metric $ds^2 = -
dt^2 + a^2(t)\big(dx^2 + dy^2 + dz^2\big).$
Investigation of (\ref{eom:1}) and finding its general solution is
a very difficult task. Hence it is important to find some special
solutions. To this end some ans\"atze of the form $\Box R
 = c R^\gamma $ seem to be useful. In the sequel we construct a few such
ans\"atze.


\subsection{Case $\Box R = r R $} \label{sec:2}

At the beginning, to illustrate method, we investigate ansatz of
the simplest form: $\Box R = r R $.  For this ansatz, where $r$ is
a constant, we have

\begin{equation}
\label{nth degree}  \Box^{n} R = r^{n}R , \,  \quad \,
\mathcal{F}(\Box) R = \mathcal{F}(r) R .
\end{equation}

In the FLRW metric $\Box = -\partial_{t}^{2} - 3 H \partial_{t}$
and ansatz $\Box R = r R$ becomes

\begin{equation} \label{equ:1}
 \ddot{R} + 3 H \dot{R} + r R = 0,
\end{equation}
where $H = \frac{\dot{a}}{a}$ is the Hubble parameter. Replacing

\begin{equation} R = 6(\dot{H} + 2 H^{2}) \label{R} \end{equation}
in  Eq. (\ref{equ:1}) we get
\begin{equation} \label{equ:2}
\dddot{H} + 4\dot{H}^{2} + 7H \ddot{H} + 12 H^{2}\dot{H} + r
(\dot{H} + 2 H^{2}) = 0.
\end{equation}
A solution of this equation is
\begin{equation}\label{H}
 H(t) = \frac{1}{2t + C_1}.
\end{equation}
This implies scale factor $a(t) = C_{2}\sqrt{ |2t + C_{1}|}$ and
acceleration $\ddot{a} =-\frac{C_2}{|2t +C_1|\sqrt{|2t+C_1|}},$
where $C_{2}>0, \, C_{1}\in \mathbb{R}.$ Calculation of $R$ by
expression (\ref{R}) gives $R = 0$ and it is consistent with other
formula containing $R,$ including (\ref{trace:1}).

It is natural to take constant $C_1 = 0,$ because it yields
symmetrical solutions with respect to $t = 0$. Result $a(t) =
C_{2}\sqrt{ |2t|}$ is an example of the symmetric singular bounce
solution.

Note that the above solutions hold also when $r = 0$ in the
ansatz, i.e.  $\Box R = 0$. More general ansatz $\Box R = r R + s$
was considered in \cite{biswas}.




\subsection{Case $\Box R = q R^2$} \label{sec:3}

The corresponding differential equation for the Hubble parameter
is
\begin{equation}
\dddot{H} + 4\dot{H}^{2} + 7H \ddot{H} + 12 H^{2}\dot{H} + 6 q (
\dot{H}^2 + 4 H^2 \dot{H} + 4 H^4) = 0
\end{equation}
with solution

\begin{equation} \label{Hn}
H_\eta(t) = \frac{2\eta+1}{3}\frac{1}{t + C_1}, \quad q_\eta
=\frac{6(\eta-1)}{(2\eta+1)(4\eta-1)},  \, \, \, \eta \in
\mathbb{R}.
\end{equation}
Another solution is $H =\frac{1}{2}\frac{1}{t + C_1}$ with
arbitrary coefficient $q$, what is equivalent to the ansatz $\Box
R = r R$ with $R = 0$.

The corresponding scalar curvature is given by
\begin{equation} \label{Rn}
R_\eta = \frac{2}{3}\frac{(2\eta+1)(4\eta-1)}{ (t+C_1)^{2}}, \, \,
\, \eta \in \mathbb{R}.
\end{equation}
It is interesting that $\Box^n R_n = 0$ when $n \in \mathbb{N}$.
This can be shown by mathematical induction by the following way.
It is evident that $\Box R_1 = 0$. Suppose that $\Box^n R_n = 0$,
then $\Box^{n+1} R_{n+1} = \Box \Box^n R_n + \frac{16n +10}{9}
\Box^n \Box R_1 = 0.$

This $\Box^n R_n = 0$ property simplifies the equations
considerably. For this special case of solutions trace equation
(\ref{trace:1}) effectively becomes
\begin{align}\label{trace:2}
 \sum_{k=1}^{n+1} f_{k} \sum_{l=0}^{k-1} (\partial_{\mu}\Box^{l}R \partial^{\mu}\Box^{k-1-l}R + 2\Box^{l}R \Box^{k-l}R )
 + 6 \Box \mathcal{F}(\Box) R = \frac{R}{8 \pi G c},
\end{align}
where
\begin{align}
\label{operatorF:n} \mathcal{F}(\Box)R = \sum_{k=0}^{n-1}
f_{k}\Box^k R.
\end{align}

In particular case $n=2$  the trace formula becomes
\begin{align} \nonumber
& \frac{36}{35}f_{0} R^{2} +  f_{1}(- \dot{R}^{2}+
\frac{12}{35}R^{3}) + f_{2}(-\frac{24}{35}R \dot{R}^{2} +
\frac{72}{1225}R^{4}) + f_{3} (-\frac{144}{1225}R^{2}\dot{R}^{2})\\
& = \frac{R}{8 \pi G c}.  \label{trace:3}
\end{align}


\subsection{Case $\Box^n R = c_n R^{\alpha n
+ \beta}$ }

We consider\footnote{B. D. thanks A. S. Koshelev for suggestion of
ansatz $\Box^n R \sim R^{n + 1}.$} another  ansatz of the form
$\Box^n R = c_n R^{\alpha n + \beta},$ where $\alpha$ and $\beta$
are constants, and $n \in \mathbb{N}$. From the  equalities
\begin{align}
&\Box^{n+1} R = \Box c_n R^{\alpha n + \beta} \nonumber\\ & = c_n
((\alpha n + \beta) R^{\alpha n + \beta - 1} \Box R - (\alpha n +
\beta) (\alpha n + \beta - 1) R^{\alpha n + \beta - 2} \dot{R}^2)
\nonumber\\& = c_n (\alpha n + \beta) (c_1 R^{\alpha n + \alpha +
2\beta -1} - (\alpha n + \beta - 1) R^{\alpha n + \beta - 2}
\dot{R}^2) = c_{n+1} R^{\alpha n + \alpha + \beta}
\label{cons:gen}
\end{align}
we get the following conditions:
\begin{align}      \label{first}
&\alpha n + \alpha + 2\beta -1 = \alpha n + \alpha + \beta, \\
&\dot{R}^2 = R^{\alpha + \beta + 1}, \label{second} \\
&c_{n+1} = c_n (\alpha n + \beta)(c_1 - \alpha n - \beta + 1).
\label{last}
\end{align}
Equation (\ref{first}) implies that $\beta$ is equal to 1.
Sequence $c_n$ is defined by (\ref{last}) and can be explicitly
written ($\beta =1$) as
\begin{equation}
c_n = c_1 \prod_{k=1}^{n-1} (\alpha k + 1)(c_1 - \alpha k),
\end{equation}
where $c_1$ is a constant. General solution of equation
(\ref{second}) is of the form
\begin{equation} \label{R:gen}
R(t)=2^{2/\alpha } \left(\alpha
   \left(\pm t - d_1\right)\right)^{-2/\alpha }, \quad  d_1 \in \mathbb{R}
\end{equation}
with arbitrary constant $d_1$.

\medskip

\noindent{\bf Case $\alpha =1.$ \, } In the case $\alpha = 1$ the
coefficients $c_n$ are given
 by $c_n = (n!)^2 {c_1 \choose n}, $ where $c_1$ is the first element.
Putting $\alpha = 1$ into equation (\ref{R:gen}) one obtains
\begin{equation} \label{R:1}
R(t) = \frac{4}{(t-d_1)^2}.
\end{equation}
 The corresponding expressions for $H(t)$
 and $a(t)$ are:
 \begin{align}
&H(t) = \frac{\left(3+\sqrt{57}\right) d_2 \left(t-d_1\right)
^{\sqrt{\frac{19}{3}}}-\sqrt{57}+3}{12 \left(t-d_1\right)
\left(d_2 \left(t-d_1\right)^{\sqrt{\frac{19}{3}}}+1\right)}, \\
& a(t)=d_3 \left(t-d_1\right)^{\frac{3-\sqrt{57}}{12}}
 \sqrt{d_2 \left(t - d_1\right)^{\sqrt{\frac{19}{3}}}+1}, \label{23}
 \end{align}
where $d_1,\, d_2,\, d_3$ are arbitrary real constants.

 \begin{figure}
\begin{center}
\includegraphics[scale=.65]{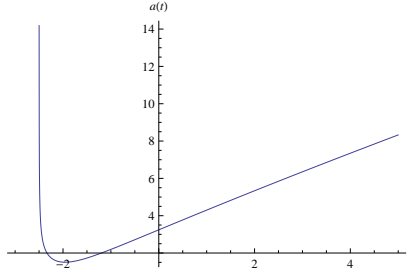}
\hskip3cm{\caption{Scale factor $a(t)$ given by (\ref{23}) for
$d_1=-2.5$, \, $d_2=2$ and $d_3=1$.}}
\label{fig:1}        
\end{center}
\end{figure}

 The function $a(t)$ has a vertical asymptote at the point $t =
 d_1$. If  $d_1<0$ then  $a(t)>0$ for all $t>0$. For
 large values of $t$,  $a(t)$ is asymptotically equivalent to
 $t^{\frac{1}{2}\sqrt{\frac{19}{3}}+\frac{1}{12} \left(3-\sqrt{57}\right)} \approx
 t^{0.879}$.
\begin{equation} \label{2dder:1}
\ddot{a}(T) = -\frac{d_3 T^{\frac{1}{12}
\left(-21-\sqrt{57}\right)} \left(\left(\sqrt{57}-5\right) d_2^2
   T^{2 \sqrt{\frac{19}{3}}}-48 d_2 T^{\sqrt{\frac{19}{3}}}-\sqrt{57}-5\right)}{24
   \left(d_2 T^{\sqrt{\frac{19}{3}}}+1\right)^{3/2}},
\end{equation}
where $T = t-d_1$. The expansion is accelerated for $d_2
(t-d_1)^{\sqrt{\frac{19}{3}}} < \frac{24}{\sqrt{57}-5}+\frac{4
\sqrt{38}}{\sqrt{57}-5}$ and it is decelerated otherwise.

Note that this ansatz $\Box^n R = c_n R^{n + 1}$ for $n=1$
coincides with ansatz $\Box R = q_\eta R^2,$ when $\eta = \frac{-1
\pm \sqrt{57}}{8},$ because then one can take $c_1 = q_\eta =
\frac{-9 \pm \sqrt{57}}{8}.$ In this particular case they have the
same scalar curvature $R$ and the same Hubble parameter for $d_2 =
0.$ However, apart from this special case $\eta = \frac{-1 \pm
\sqrt{57}}{8},$ constant $q_\eta$ is different of $c_1 = \frac{-9
\pm \sqrt{57}}{8}.$

\medskip

\noindent{\bf Case $\alpha = \frac{1}{2}$. \, } Putting $\alpha =
\frac{1}{2}$ into equation (\ref{R:gen}) we obtain
\begin{equation} \label{R:1/2}
R(t)= \frac{256}{\left(t+d_1\right){}^4}.
\end{equation}
From (\ref{R}) we obtain
\begin{equation}
H(t) = \frac{-512 \sqrt{3} d_2 e^{\frac{32}{\sqrt{3}
   \left(d_1+t\right)}}+ 3 (t +d_1) \left(32 d_2 e^{\frac{32}{\sqrt{3}
   \left(d_1+t\right)}}+\sqrt{3}\right)+48}{6
   \left(d_1+t\right){}^2 \left(32 d_2 e^{\frac{32}{\sqrt{3}
   \left(d_1+t\right)}}+\sqrt{3}\right)}
\end{equation}
and then
\begin{equation}
a(t)= d_3 e^{-\frac{8}{\sqrt{3} \left(d_1+t\right)}} \sqrt{d_1+t}
   \sqrt{32 d_2 e^{\frac{32}{\sqrt{3}
   \left(d_1+t\right)}}+\sqrt{3}}, \label{27}
 \end{equation}
where $d_1,\, d_2,\, d_3$ are some real constants.

 \begin{figure}
 \begin{center}
\includegraphics[scale=.65]{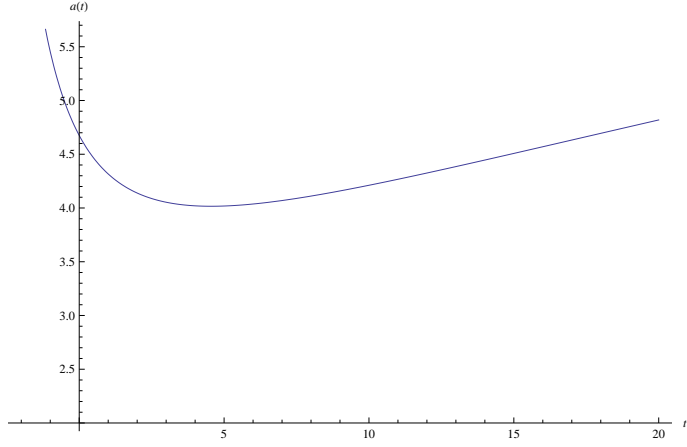}.
\caption{Scale factor $a(t)$ given by (\ref{27}) for
$d_1=\frac{8}{\sqrt 3}$, $d_2=2$ and $d_3=\frac{1}{10}$.}
\end{center}
\label{fig:2}       
\end{figure}


{The corresponding acceleration is
\begin{align}
\ddot{a}(t)&=\frac{d_3 e^{-\frac{8}{\sqrt{3}
\left(d_1+t\right)}}}{12
   \left(d_1+t\right){}^{7/2} \left(32 d_2 e^{\frac{32}{\sqrt{3}
   \left(d_1+t\right)}}+\sqrt{3}\right){}^{3/2}}
   \Big( 1024 \,d_2^2 \, e^{\frac{64}{\sqrt{3} \left(c_1+t\right)}} \nonumber \\ &\times\left(-6 d_1
   t-3 d_1^2+32 \sqrt{3} d_1-3 t^2+32 \sqrt{3} t+256\right) \nonumber \\
   &-3 \left(6 d_1 t+3 d_1^2+32 \sqrt{3} d_1+3 t^2+32 \sqrt{3}
   t-256\right) \nonumber \\
   &-192 \sqrt{3} d_2 e^{\frac{32}{\sqrt{3} \left(d_1+t\right)}} \left(d_1+t-16\right)
   \left(d_1+t+16\right) \Big).
\end{align}}
The acceleration is positive for $t < t_0$ and negative for $t >
t_0,$ where $t_0$ is the zero of $\ddot{a}(t).$  For large values
of $t$,  $\ddot{a}(t)$ converges to 0.

\medskip



\noindent{\bf Case $\alpha = 2$. \, } In  the case $\alpha = 2$
only one integration can be performed and it gives
\begin{align}
&H(t)=\frac{
 -\sqrt{3} d_2 I_1\left(\frac{2 \sqrt{t-d_1}}{\sqrt{3}}\right)
 -d_2 \sqrt{t-d_1} I_0\left(\frac{2 \sqrt{t-d_1}}{\sqrt{3}}\right)
 -d_2 \sqrt{t-d_1} I_2\left(\frac{2 \sqrt{t-d_1}}{\sqrt{3}}\right)
}{4 \sqrt{3} \left(t-d_1\right)
   \left(K_1\left(\frac{2 \sqrt{t-d_1}}{\sqrt{3}}\right)-d_2
   I_1\left(\frac{2 \sqrt{t-d_1}}{\sqrt{3}}\right)\right)}
   \nonumber \\
&+ \frac{
 -\sqrt{t-d_1} K_0\left(\frac{2 \sqrt{t-d_1}}{\sqrt{3}}\right)
 +\sqrt{3} K_1\left(\frac{2 \sqrt{t-d_1}}{\sqrt{3}}\right)
 -\sqrt{t-d_1} K_2\left(\frac{2 \sqrt{t-d_1}}{\sqrt{3}}\right)
}{4 \sqrt{3} \left(t-d_1\right)
   \left(K_1\left(\frac{2 \sqrt{t-d_1}}{\sqrt{3}}\right)-d_2
   I_1\left(\frac{2 \sqrt{t-d_1}}{\sqrt{3}}\right)\right)}.
\end{align}
$I_i$ and $K_i$ are modified Bessel functions of the first and the
second kind, respectively, and $d_1$, $d_2$ are real constants.

\medskip

\noindent{\bf Case $\alpha = -2$. \, } For $\alpha=-2$ we  obtain
expression for $H(t)$ involving Airy functions
\begin{eqnarray}
H(t)=\frac{\sqrt[3]{-\frac{1}{3}} \left(d_2
   \text{Ai}'\left(\sqrt[3]{-\frac{1}{3}}
   \left(t-d_1\right)\right)+\text{Bi}'\left(\sqrt[3]{-\frac{1}{3}}
   \left(t-d_1\right)\right)\right)}{2 \left(d_2
   \text{Ai}\left(\sqrt[3]{-\frac{1}{3}}
   \left(t-d_1\right)\right)+\text{Bi}\left(\sqrt[3]{-\frac{1}{3}}
   \left(t-d_1\right)\right)\right)}.
\end{eqnarray}

\section{Concluding Remarks}

In this article we presented three ans\"atze, two of them are
quite new and one can be adjusted so that $\Box^n R =0$. These two
ans\"atze have solutions for scalar curvature of the form $R =
\frac{C_2}{(t + C_1)^2}$, which satisfy all but  extended Einstein
equations (\ref{eom:1}) and related trace formula (\ref{trace:1}).
It is a consequence of the quadratic form in $R$ of the Lagrangian
(\ref{lag:1}). However these ans\"atze are  promising for some new
nonlocal Lagrangians, which investigation is in progress.

It is worth mentioning that all the above ans\"atze contain
solution $R =0$, which satisfies all (including (\ref{eom:1}) and
(\ref{trace:1})) equations with curvature constant $k= -1.$
Namely, for $R=0,$ Eq. (\ref{eom:1}) reduces to $G_{\mu\nu} = 0$
and it gives
\begin{align} \label{9.03-1}
\frac{\ddot{a}}{a} = 0 , \quad \quad \Big(\frac{\dot{a}}{a}\Big)^2
+ \frac{k}{a^2} = 0.
\end{align}
If $k = 0$ one has only static  solution $a = constant.$ However,
when $k = - 1$ then $a(t) = |t|$ and it contains a crunch
preceding to a big bang.

Above considered ans\"atze may be also useful in analysis of some
other nonlocal gravity and cosmology models. Further investigation
of nonlocality governed by the  Riemann zeta function in $p$-adic
strings dynamics \cite{dragovich} extends interesting cases and
can give new insights.

\section*{Acknowledgements}
This investigation is supported by Ministry of Education and
Science of the Republic of Serbia, grant No 174012. B. D. is
grateful to Alexey Koshelev for useful discussions. Authors thank
anonymous referee for constructive comments to improve
presentation and clarify some assertions. This is an extended
version of talk presented at the IX International Workshop ``Lie
Theory and its Applications in Physics'', 20--26 June 2011, Varna,
Bulgaria and I. D. thanks organizers for hospitality.

\end{document}